\documentclass{ragtime}
\usepackage[utf8]{inputenc}
\usepackage[T1]{fontenc}
\usepackage{graphicx}
\usepackage{wrapfig}
\usepackage{amsmath}
\usepackage{placeins}
\usepackage{graphicx}

\usepackage{hyperref}
\hypersetup{
    colorlinks=true,       % Colors the text instead of surrounding boxes
    linkcolor=blue,        % Color for internal links (sections, pages)
    citecolor=blue,       % Color for bibliographic citations
    urlcolor=blue,      % Color for external URLs
    filecolor=blue,     % Color for local file links
    pdfpagemode=UseBookmarks,
    pdftitle={Your Document Title}
}

\title[NS-ULXs accretion columns]%
{Super-Eddington Neutron Star Accretion Columns: Comparing GRRMHD Simulations with Spectral Models}
\author[Fatemeh Kayanikhoo, Marios Kouzis]% running-head authors
{Fatemeh Kayanikhoo\at{1}
Marios Kouzis\at{2}\\
\ins{1}Research Centre for Computational Physics and Data Processing\\
in Physics Institute, Silesian University in Opava\\
\ins{2}Nicolaus Copernicus Astronomical Center, Polish Academy of Sciences}%

\makeatletter
\let\NaviBar\@empty
\def\ps@titlepage{\ps@headings\def\@oddhead{\hfill\thepage}\let\@evenhead\@oddhead\let\@oddfoot\@empty\let\@evenfoot\@empty}
\makeatother

\begin{document}

\begin{abstract}

The geometry and structure of neutron star accretion columns affect the X-ray emission in super-Eddington accreting systems. We perform four axisymmetric general relativistic radiation magnetohydrodynamic simulations of super-Eddington accretion onto magnetized neutron stars with the maximum surface magnetic fields ranging from $3\times10^{10}$ to $1\times10^{11}\,\mathrm{G}$. We measure the column width and estimate the characteristic radiation shock heights from the flow velocity and velocity gradient. Across the four models, the angular width decreases from approximately $0.14$ to $0.07\,\mathrm{rad}$, while the estimated shock height increases from $1.95$ to $3.89\,\mathrm{km}$. We then compare our solutions to the best-fit values derived from spectral modeling of super-Eddington Be/X-ray binary outbursts. We calculate area-equivalent column radii that vary between 5.2 and 3.2 km, three of which fall within the upper part of the range inferred from spectral results using the \texttt{bwcycl} model.
\end{abstract}

\begin{keywords}
neutron stars~-- accretion columns~-- radiation magnetohydrodynamics~-- general relativity~--
ultraluminous X-ray sources~-- radiation shocks
\end{keywords}

\section{Introduction}\label{sec:intro}

The detection of coherent pulsations from M82 X-2 consistent with the neutron star (NS) spin period established
that accreting NSs can power ultraluminous
X-ray sources (ULXs), with an isotropic luminosity 100--1000 times above their Eddington limit \citep{Bachetti2014}. The critical luminosity of ULXs was explained by beaming proposed by \citet{king2001} and further studied by KLK model \citep{king2017,king2019,king2020}. Numerical simulations have further demonstrated that super-Eddington accretion onto magnetized NSs can produce beamed emission consistent with those observed in ULXs \citep{abarca+21,Kayanikhoo2025a,Kayanikhoo2025b}.

A sufficiently strong magnetic field channels matter toward the magnetic poles, where radiation pressure can decelerate the inflow above the stellar surface and support a dense settling region, so-called radiation-mediated shock. The geometry of this region influences how radiation escapes from the accretion flow. In the classical picture, at subcritical accretion, X-ray emission is directed preferentially along the magnetic axis, ``pencil-beam'', whereas at supercritical accretion, X-ray emission escapes mainly through the side walls
of an extended accretion column, ``fan-beam'' \citep{BaskoSunyaev1976}. Luminosity-dependent changes in the pulse profiles of \textit{Swift J0243.6+6124} have been interpreted in terms of changes in these emission geometries
\citep{WilsonHodge2018,Liu2022}.

The structure of an accretion column also affects its X-ray spectrum. The velocity and density distributions influence bulk and thermal Comptonization, while the column geometry affects the paths through which photons escape
\citep{BeckerWolff2022}. Using a radiation hydrodynamic accretion column model, \citet{West2024} reproduced the spectra of \textit{RX J0209.6$-$7427} during two luminosity states, obtaining an outer polar-cap radius of approximately
$0.64\,\mathrm{km}$ and radiation sonic-point heights of $4.9$--$5.3\,\mathrm{km}$ above the NS surface.

Cyclotron resonance scattering features provide another observational constraint on the geometry of accretion flow. In \textit{V 0332+53}, the cyclotron line energy decreases with increasing X-ray luminosity \citep{Tsygankov2006}. One interpretation is
that the line-forming region moves farther above the stellar surface as the luminosity increases, where the local magnetic field is weaker. Such observations suggest that the vertical structure of the accretion column changes with luminosity. These observations motivate numerical studies of accretion column geometry and radiation shock structure.

Published numerical studies have investigated supercritical accretion columns using different geometries and physical assumptions. \citet{Kawashima2016} performed two-dimensional radiation hydrodynamic simulations and showed that radiation pressure supports dense settling regions above the stellar surface, enabling photons to escape laterally. At substantially higher accretion rates, similar calculations by  \citet{KawashimaOhsuga2020} produced hollow-cone accretion structures. More recent relativistic radiation magnetohydrodynamics (MHD)
simulations examined the effects of split-monopole magnetic field configuration and scattering opacity on column structure and its variability \citep{Sheng2023}. A subsequent parameter survey extended these calculations to different accretion rates, magnetic
fields, and flow geometries \citep{Zhang2025}.

We present four general relativistic radiative MHD (GRRMHD) simulations of accretion onto NSs with a surface dipolar magnetic field with strengths ranging from $3\times 10^{10}$ to $1\times 10^{11}\,$G. We measure the accretion column geometry and estimate the shock height. Then we compare the equivalent parameters to those from observational models.

This paper is structured as follows: Section~\ref{sec:method} presents numerical methods and simulation details. In Section~\ref{sec:geom} we compute the geometry of the accretion column. In Section~\ref{sec:rad}, the radiation shock height is computed, and Section~\ref{sec:conclusions} compares the model parameters with those of the observational models and summarizes the main conclusions.

\section{Numerical methods and simulation setup}\label{sec:method}

The simulations are performed using the GRRMHD code \textsc{Koral}, developed by \citet{Sadowski2013}, with the NS implementation by \citet{abarca+21}. The code evolves the gas and radiation in a fixed spacetime by solving the mass and energy-momentum conservation equations,

\begin{equation}
\nabla_\mu(\rho u^\mu)=0,\quad
\nabla_\mu T^\mu{}_\nu=G_\nu,\quad
\nabla_\mu R^\mu{}_\nu=-G_\nu,
\label{eq:grrmhd}
\end{equation}

\noindent where $\rho$ is the rest-mass density, $u^\mu$ is the fluid four-velocity, $T^\mu{}_\nu$ and $R^\mu{}_\nu$ are the MHD and radiation stress-energy tensors, respectively, and $G_\nu$ is the radiation four-force describing energy and momentum exchange between matter and radiation. 

Radiation is treated using the M1 closure scheme. The radiation-matter interaction includes electron scattering, free-free emission and absorption, and Comptonization. The low-density, strongly magnetized region within the magnetosphere approaches the force-free regime developed by \cite{abarca+21}. The stellar magnetic field is initialized using the relativistic dipole potential \citep{Wasserman+1983}. The magnetic field is subsequently evolved through the MHD induction equation using the constrained transport method.

\subsection*{Simulation setup} We adopt a NS mass of $M_\star=1.4\,M_\odot$ and a stellar radius of $R_\star=5\,r_{\rm g}\simeq10.34\,\mathrm{km}$, where $r_{\rm g}=GM_\star/c^2$. The simulations are initialized with an equilibrium torus with a single-loop magnetic field, extending radially from $34\,r_{\rm g}$ to $500\,r_{\rm g}$ \citep[see][ and references therein]{Kayanikhoo2025b}. We perform the simulations in Schwarzschild spacetime on an axisymmetric (2.5D) grid with a resolution of $N_r\times N_\theta=512\times510$. The radial grid is logarithmically spaced, while the polar angle is uniformly spaced from $0.002$ to $(\pi-0.002)$\footnote{We exclude a small angle near the poles to avoid numerical issues.}. The region extending to $5\,r_{\rm g}$ above the stellar surface in each hemisphere contains approximately $67\times255$ grid cells. We set this resolution to adequately resolve the accretion column, the magnetic field structure near the inner boundary, and the relevant flow features while minimizing numerical errors. 

At the stellar surface, we impose an absorbing boundary condition for the accreting matter, following \citet{ParfreyTchekhovskoy2017}. We use the radiation-reflective boundary surface as described in \citet{abarca+21}, with a prescribed surface albedo of $0.75$.

In this paper, we present the simulations that are labeled as \texttt{B03}, \texttt{B05}, \texttt{B07}, and \texttt{B10}, each number indicates the maximum surface magnetic field strength, $B_\star$, in units of $10^{10}\,\mathrm{G}$. Each simulation ran for $50,000 \, t_{\rm g}$ (where the gravitational time is $t_{\rm g} = GM/c^3$), equivalent to a physical time of about 0.35 seconds. The reported results are from time-averaged data over an interval of about $35,000\, t_{\rm g}$. We show only the northern hemisphere. However, when calculating the column parameters, we use the average values from the northern and southern accretion funnels. 

\section{Angular geometry of accretion column}\label{sec:geom}

\begin{figure}[ht!]
    \centering
    \includegraphics[width=1\linewidth]{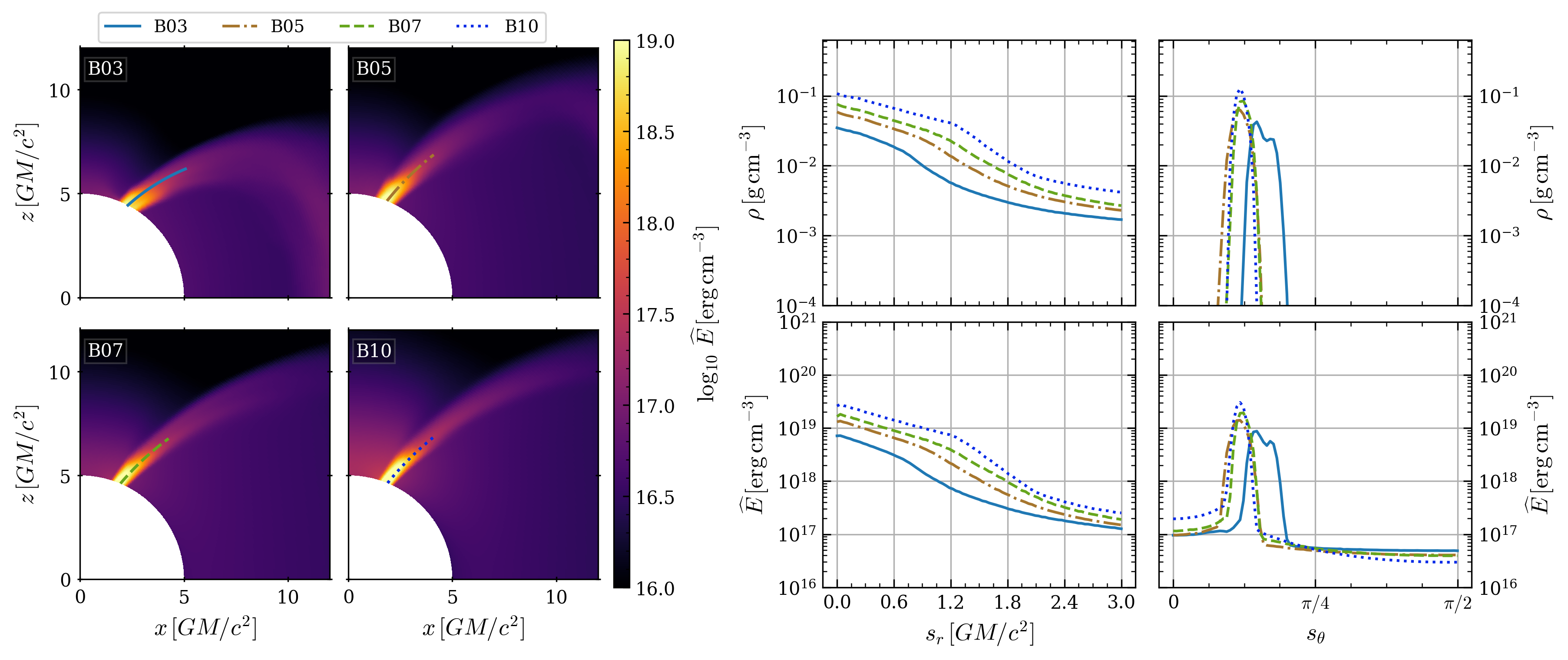}
    \caption{\textit{Left:} colourmaps show logarithmic values of radiation energy density $\hat{E}$; curves on each panel indicate the traced paths through the one-pole accretion funnel. \textit{Right:} The radial and angular profiles of rest-mass density $\rho$ (upper panels) and $\hat{E}$ (lower panels). The middle panels show quantities along the traced paths as functions of $s_r$, with the initial point on the surface of the NS. The right panels show angular profiles of quantities at a fixed radius immediately above the stellar surface, shown as functions of $s_\theta$.}
    \label{fig:column}
\end{figure}

To characterize the accretion column structure, we examine the comoving radiation energy density $\hat{E}$ and the gas rest-mass density $\rho$ as shown in Fig.~\ref{fig:column}. For each simulation, we locate the maximum gas density near the stellar surface and trace a path outward along the poloidal magnetic field. The traced paths are shown as lines in the colormap panels.
We then interpolate $\rho$ and $\hat{E}$ along these paths to obtain the profiles shown in the middle panels. Here, $s_r$ is the cumulative path coordinate, measured outward from the stellar surface. 

Both $\rho$ and $\hat{E}$ increase toward the NS surface. At each radius above the stellar surface, their values increase with the strength of the NS surface magnetic field. We note that although all simulations have similar initial conditions and aim for an accretion rate $\dot{M}$ of about 20 times the Eddington limit, the resulting accretion rates increase with magnetic field strength \footnote{The accretion rate is measured at approximately $20\,r_{\rm g}$, outside the magnetosphere, and where the disk has reached a quasi-steady state \citep[see][ for more details]{Kayanikhoo2025b}.} (see $\dot{M}$ in Table~\ref{tab:column_geometry}).
The gradients of $\rho$ and $\hat{E}$ vary along the traced paths; however, neither quantity shows a clear discontinuity that would allow us to locate the radiation shock height directly from these profiles. Similar increases in $\rho$ and $\hat{E}$ toward the stellar surface within the post-shock settling region have been reported in previous radiation-MHD simulations \citep{Zhang2023,Zhang2025}. We therefore use the radial velocity and its gradient to estimate the shock height, as described in the following section.

To examine the angular structure of the column, we extract $\rho(r_0,s_\theta)$ and $\hat{E}(r_0,s_\theta)$ along a fixed-radius arc just above the stellar surface, where $r_0 \simeq R_\star$ and $0\lesssim s_\theta\leq \pi/2$. The right-hand panels of Fig.~\ref{fig:column} show these angular profiles as functions of $s_\theta$.
We determine the column position and width from the angular profile of $\hat{E}$. First, we estimate and subtract the background radiation level using values near both ends of the profile. We define the column position $\theta_{\rm c}$ as the location of the strongest peak in the background-subtracted profile. The angular width $w_\theta$ is defined as the full width at half maximum (FWHM) of this peak. 

The measured column parameters are summarized in Table~\ref{tab:column_geometry}. As the strength of the NS surface magnetic field increases from model \texttt{B03} to \texttt{B10},
$w_\theta$ decreases from $0.14$ to $0.07\,\mathrm{rad}$. The peak position $\theta_{\rm c}$ varies between $0.46$ and $0.37\,\mathrm{rad}$, with the strongest magnetic field model showing a smaller angular distance from the magnetic axis.

We also calculate the area-equivalent cylindrical radius $r_{0,\rm eq}$ of the accretion column on the NS surface for comparison with observational models. The results are listed in Table~\ref{tab:column_geometry} and discussed in Section~\ref{sec:conclusions}.

\begin{table}[t!]
\centering
\caption{Accretion rate $\dot{M}$, the column position
$\theta_{\rm c}$, measured from the magnetic axis,
and $w_\theta$, the angular width of the radiation energy
profile. $r_{0,\rm eq}$ is the area-equivalent cylinder radius.}
\label{tab:column_geometry}
\resizebox{0.80\linewidth}{!}{%
\begin{tabular}{lccccc}
\hline
Model
& $B_\star$ 
& $\dot M$ 
& $\theta_{\rm c}$ 
& $w_\theta$
& $r_{0,\rm eq}$  \\
& ($10^{10}\,G$)

& ($10^{19}\,\mathrm{g\,s^{-1}}$)
& (rad)
& (rad)
& (km)\\
\hline
B03  & 3  & 6.72 & 0.462 & 0.141 & 5.18 \\
B05  & 5  & 6.91 & 0.370 & 0.099 & 3.91 \\
B07 & 7 & 8.37 & 0.388 & 0.078 & 3.55 \\
B10 & 10 & 9.54 & 0.370 & 0.067 & 3.22 \\
\hline
\end{tabular}%
}
\end{table}

\section{Radiation shock}\label{sec:rad}

We estimate the radiation shock height from the inward radial velocity profile $v_{\rm r}$ along each traced accretion column. The inward velocity is calculated as

\begin{equation}
v_{\rm r}=-\frac{u^r}{u^t}
\sqrt{\frac{g_{rr}}{-g_{tt}}},
\end{equation}

\noindent where $u^\mu$ is the gas four-velocity and $g_{\mu\nu}$ is the Schwarzschild metric. The velocity is interpolated along each traced path and plotted as a function of $s_r$ in km units.

\begin{figure}[t]
    \centering
    \includegraphics[width=\linewidth]{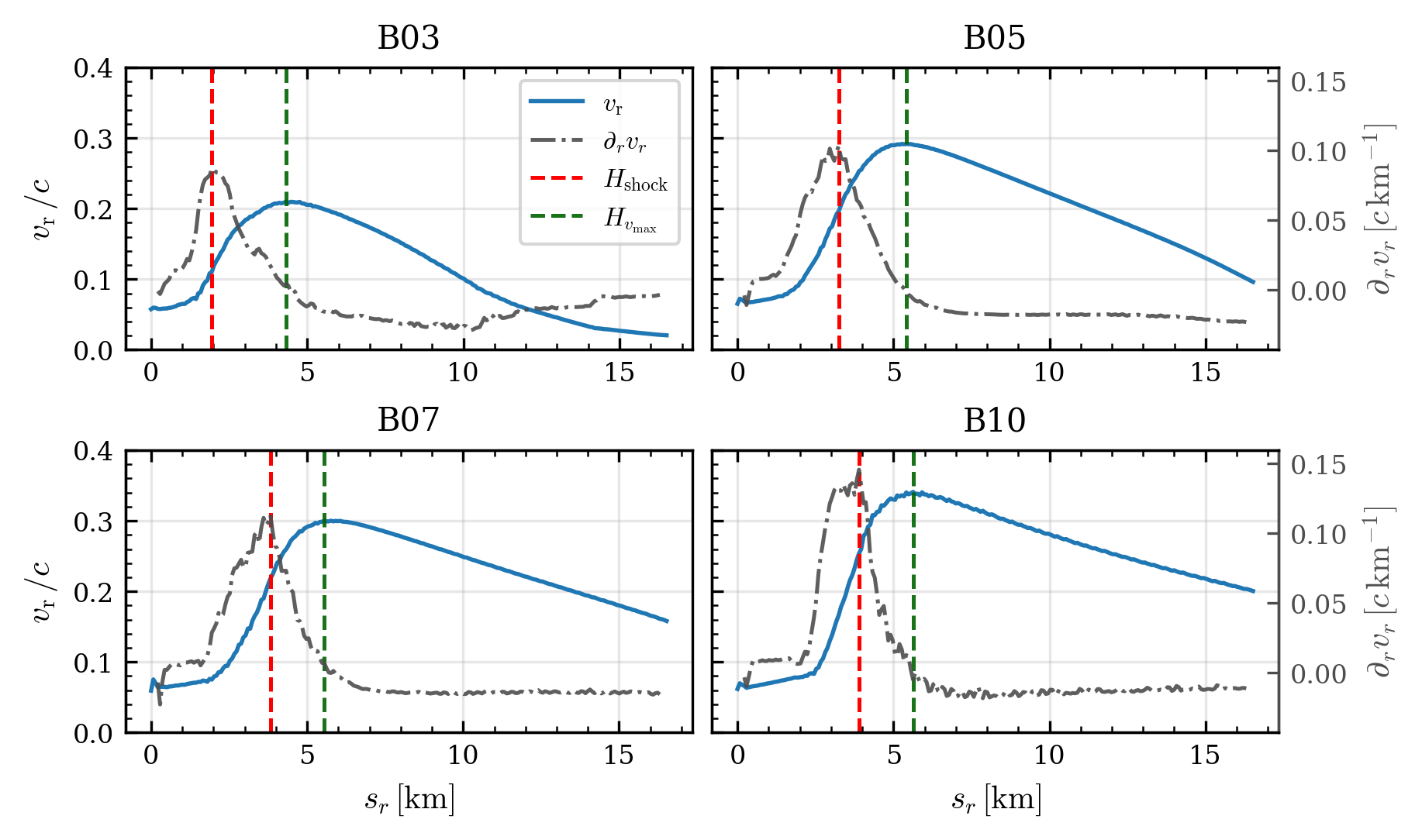}
    \caption{Radial profile of the inward radial velocity $v_{\rm r}$ (blue solid lines)
and radial velocity gradients $\partial_r v_r$ (grey dash-dotted lines)
along the traced accretion columns. The gradients are calculated from velocity profiles.
The red dashed lines indicate the estimated radiation shock
heights, $H_{\rm shock}$, identified by the maximum positive $\partial_r v_r$. The green dashed lines mark the heights of maximum inward velocity, $H_{v_{\max}}$.}
    \label{fig:velocity}
\end{figure}

The inward radial velocity profile is smoothed using a five-point moving average, and its gradient $ \ partial_r v_r$ is calculated with respect to the radial coordinate along the traced path. The shock height $H_{\rm shock}$ is estimated from the location of the maximum positive velocity gradient above the stellar surface, corresponding to the region of strongest deceleration. 
We also determine the height of maximum inward velocity $H_{v_{\max}}$ along each traced path (see Fig.~\ref{fig:velocity}). 

We estimate the sonic-point height $H_{\mathcal{M}}$ from the location where the radiation Mach number 
reaches unity. The density-weighted mean radial Mach number $\langle\mathcal{M}_r\rangle_\rho$ is computed as 

\begin{equation}
\langle\mathcal{M}_r\rangle_\rho(r)
=
\frac{
\int_{\theta_1}^{\theta_2}
\rho\,\mathcal{M}_r\sin\theta\,d\theta
}{
\int_{\theta_1}^{\theta_2}
\rho\sin\theta\,d\theta
},
\end{equation}

\noindent where $\mathcal{M}_r=|v_r|/a_{\rm rad}$,
with the radiation sound speed
$a_{\rm rad}=2/3(\hat{E}/\rho)^{1/2}$. 
The angular integration is restricted to the dense region
of the accretion column.

The estimated shock heights $H_{\rm shock}$ and $H_{v_{\rm max}}$ and their radial separations $\Delta H$, the maximum inward velocity $v_{\rm{max}}$ along with $H_\mathcal{M}$ are summarized in
Table~\ref{tab:shock}.

\begin{table}[t]
\centering
\caption{Estimated radiation shock heights $H_{\rm shock}$,
heights of the
velocity maxima $H_{v_{\max}}$, radial separations
$\Delta H=H_{v_{\max}}-H_{\rm shock}$, maximum inward velocities $v_{\max}$, sonic-point height $H_{\mathcal{M}}$ and mean
lateral luminosities $\overline{L_{\rm side}}$
for each model.}
\label{tab:shock}
\resizebox{0.80\linewidth}{!}{%
\begin{tabular}{lcccccc}
\hline
Model
& $H_{\rm shock}$
& $H_{v_{\max}}$
& $\Delta H$
& $v_{\max}$
& $H_{\mathcal{M}}$
& $\overline{L_{\rm side}}$ \\
& (km)
& (km)
& (km)
& ($c$)
& (km)
& ($10^{38}\,\mathrm{erg\,s^{-1}}$) \\
\hline
B03  & 1.95  & 4.32 & 2.37 & 0.21&3.25 & 4.46 \\
B05  & 3.25 & 5.41 & 2.16 & 0.29& 3.26&3.89  \\
B07 & 3.82  & 5.55 & 1.73 & 0.30& 3.72&3.17 \\
B10 & 3.89  & 5.63 & 1.74 & 0.34& 3.91&2.01 \\
\hline
\end{tabular}%
}
\end{table}

Our results show that the estimated heights $H_{\rm shock}$ and
$H_{v_{\max}}$ and the sonic-point height $H_{\mathcal{M}}$ increase from model \texttt{B03} to \texttt{B10}. In all four models, the estimated $H_{\rm shock}$ is smaller than $H_{v_{\max}}$ but nearly equal to the sonic-point height, $H_{\mathcal{M}}$, except in  \texttt{B03} model. 
The maximum inward velocity also increases, with a
larger relative change than its height. Meanwhile,
the separation $\Delta H$ decreases from 2.37 to 1.74~km. These results suggest that the inflow reaches higher velocities farther above the stellar surface, while the region of strongest deceleration shifts upward and lies closer to the velocity maximum in the stronger magnetic field models.

We measure the luminosity escaping from the photosphere \citep[see equation~18 of][]{Kayanikhoo2025b} adjacent to the lateral sides of the accretion columns. The luminosities from the northern and southern regions,
$L_{\rm N}$ and $L_{\rm S}$, respectively, are calculated
within $10\,\mathrm{km}$ above the NS surface.
We define the mean side luminosity as $\overline{L_{\rm side}}=(L_{\rm N}+L_{\rm S})/2$. Across the four models, the mean lateral radiation luminosity decreases with increasing stellar magnetic field strength, even though the mass accretion rate also increases (see Table~\ref{tab:shock}. This may suggest stronger radiation advection within the denser, more strongly magnetized accretion columns. 
We note that $\overline{L_{\rm side}}$ is neither the isotropic luminosity nor the total bolometric luminosity of the source. The luminosities of these simulations are discussed in detail in \cite{Kayanikhoo2025b}.

\section{Discussion and conclusions}\label{sec:conclusions}

% \subsection{Connection to observational models}
A close observational analogue of ULX pulsars may be represented by major Type II outbursts of Be/X-ray binaries (BeXRBs). During these rare events, the bolometric X-ray luminosity of these systems can exceed the Eddington limit for a canonical NS ($\sim 2 \times 10^{38} \  \text{erg} \ \text{s}^{-1}$) by a factor of 10 or more, and in most extreme cases it reaches ULX levels. The Galactic BeXRB \textit{Swift J0243.6+6124} was identified as the first Galactic ULX pulsar during its 2017-2018 giant outburst \citep{WilsonHodge2018}, \textit{SMC X-3} reached $\sim2.5 \times 10^{39}\  \text{erg} \ \text{s}^{-1}$ in 2016-2017 \citep{Tsygankov2017}, and \textit{RX J0209.6-7427} reached $\sim2 \times 10^{39}\  \text{erg} \ \text{s}^{-1}$ in 2019 \citep{Vasilopoulos2020}. Sources in the Magellanic Clouds are particularly valuable because their well-defined distances and low foreground absorption allow their luminosities to be measured far more reliably than for Galactic systems, and several of their super-Eddington outbursts have been covered by \textit{NuSTAR}. Therefore, major BeXRB outbursts are ideal candidates for understanding super-Eddington accretion by the application of spectral models.

The analytical model of \citet[][, hereafter BW07]{BW07} computes the thermal and bulk Comptonization of blackbody, cyclotron and bremsstrahlung seed photons in a cylindrical, radiation-dominated accretion column of radius $r_0$ with a constant magnetic field, and is available in \textsc{xspec} as \texttt{bwcycl} \citep{Ferrigno2009}. \citet{Vasilopoulos2025} applied this model to super-Eddington outbursts in the Magellanic Clouds, and Kouzis et al.\ (in prep.) extend the analysis to 13 \textit{NuSTAR} observations of five BeXRBs (\textit{RX J0520.5-6932, SMC X-2, RX J0209.6-7427, SXP 31, SXP 4.78}) whose outbursts peaked at $L_{\rm X} \approx (1.8$--$20) \times 10^{38} \  \text{erg} \ \text{s}^{-1}$. The best spectral fits require $\dot M\approx(0.9$--$4.3)\times10^{18}$\,g\,s$^{-1}$,
$B\approx(3$--$10)\times10^{12}$\,G and $kT_{\rm e}\approx3$--$7$\,keV, and emission from Comptonized bremsstrahlung photons. They also require column radii $r_0 \approx 0.8$--$4.9$ km, about an order of magnitude larger than the values typically obtained from sub-Eddington pulsars. In BW07 the top of the column, $z_{\max}$, is the height at which the model velocity profile equals the local free-fall speed, i.e. where the inflow begins to decelerate. For the fitted sources, $z_{\max}$ ranges from $\approx2$--$15\ R_\star$ and increases with $L_{\rm X}/L_{\rm Edd}$.

In BW07, the accreting material is channeled through a cylinder of radius $r_0$ and in order to compare the geometry of our simulations with the one inferred from best-fit results, we define an equivalent cylinder that carries the same mass accretion rate at the same density and velocity. Because our simulations are axisymmetric, the accretion column can be considered as a hollow cone that intersects with the stellar surface through a ring of angular width $w_\theta$ centered at $\theta_{\rm c}$. Therefore integrating the surface element $dA= R_{\star}^2 \sin \theta d \theta d \phi$ over the ring gives

\begin{equation}
    A_{\text{ring}} = 2 \pi R_{\star}^2 \left[\cos(\theta_{\rm c} - w_{\theta}/2) - \cos(\theta_{\rm c} + w_{\theta}/2)\right]
    \label{eq:Aring}
\end{equation}

The BW07 solutions assume a single cylinder, whereas in the simulations $\dot M$ is shared between the two magnetic poles. The equivalent cylinder must therefore have the combined area of both rings, $\pi r_{0,{\rm eq}}^2=2A_{\rm ring}$, which gives

\begin{equation}
    r_{0,{\rm eq}}= \sqrt{\frac{2A_{\rm ring}}{\pi}} = 2R_{\star} \left[ \cos(\theta_{\rm c} - w_{\theta}/2) - \cos(\theta_{\rm c} + w_{\theta}/2)\right]^{1/2}
    \label{eq:r0eq}
\end{equation}

For our four models, $r_{0,{\rm eq}}\approx5.2$, 3.9, 3.6 and 3.2\,km (see Table~\ref{tab:column_geometry}). These values overlap with the upper part of the range obtained from the \texttt{bwcycl} fits ($r_0\approx0.8$--4.9\,km). The large radii inferred from the spectra may therefore partly reflect two thin annular columns described by a filled cylinder.

\begin{figure}[h!]
    \centering
    \includegraphics[width=0.5\linewidth]{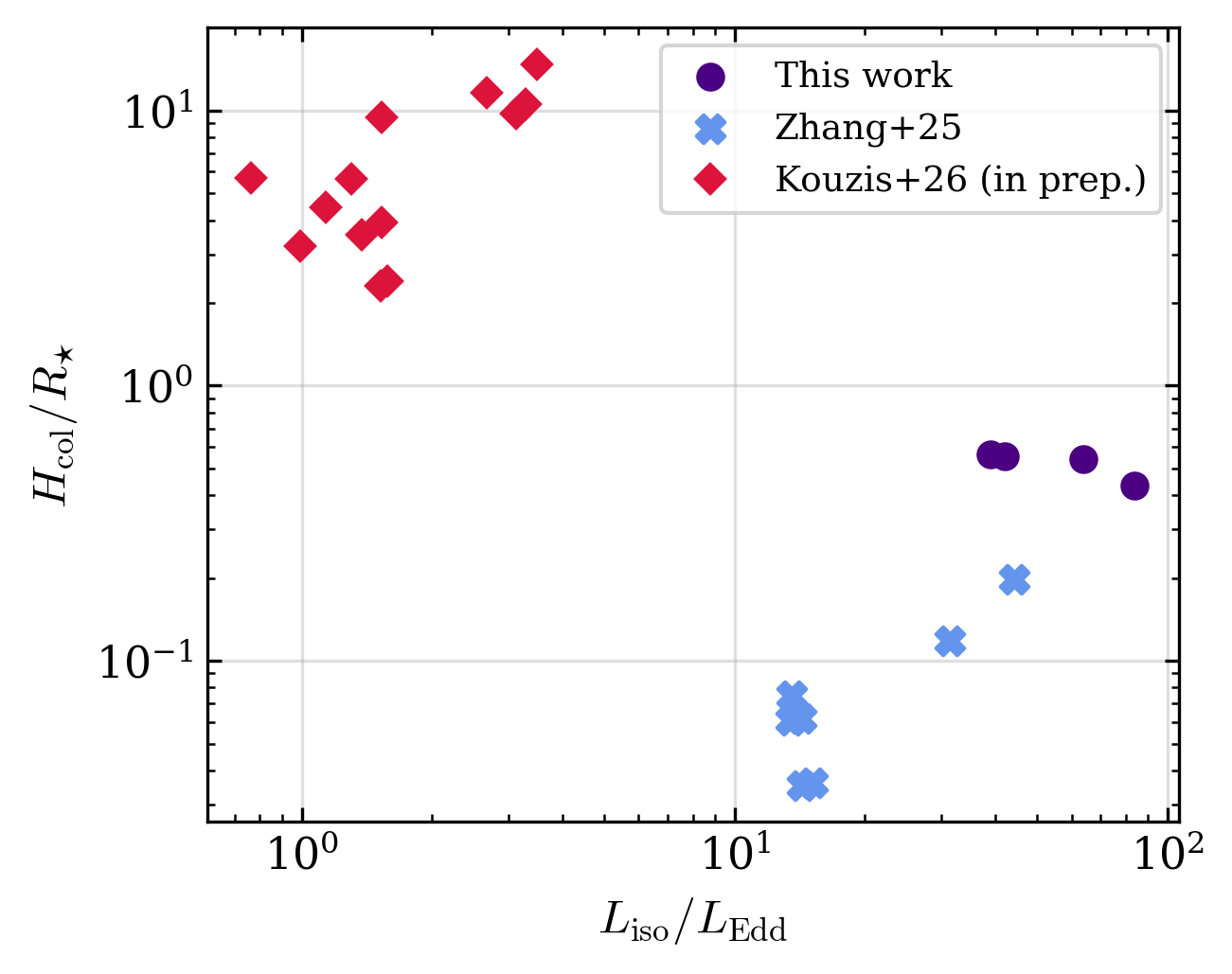}
    \caption{Accretion-column height in units of the NS radius versus apparent luminosity in Eddington units. Red diamonds: $z_{\max}$ from \texttt{bwcycl} fits to 13 \textit{NuSTAR} observations of five BeXRBs in the Magellanic Clouds during super-Eddington outbursts (Kouzis et al., in prep.). Purple circles: height of maximum inflow velocity, $H_{v_{\max}}$, in our simulations. Blue crosses: shock heights from the radiation-MHD simulations of \citet{Zhang2025}.}
    \label{fig:hmax}
\end{figure}

The column heights estimated from our simulations agree less well with those of \texttt{bwcycl}. In order to make a close comparison with the best-fit $z_{\max}$, we use $H_{v_{\max}}$, the height at which the inflow starts to decelerate. The values from our simulations, 0.42--0.54\,$R_\star$, lie well below the \texttt{bwcycl} heights of $z_{\max}\gtrsim2\,R_\star$ (Fig.~\ref{fig:hmax}). Part of the difference between the simulated and fitted heights may have to do with the upper boundary condition of BW07. In our simulations, the inflow never reaches free fall. In BW07, the velocity grows only as $z^{1/2}$ above the stellar surface, so requiring it to match the free-fall speed places the top of the column at large heights. On the other hand, the heights at which our inflow becomes subsonic, $\approx 3.2$--$3.9$~km (see $H_{\mathcal{M}}$ in Table~\ref{tab:shock}), are comparable to the radiation sonic-point heights obtained for \textit{RX J0209.6-7427} with the self-consistent radiation-hydrodynamic model of \citet{West2024}.

These comparisons are preliminary, because the simulated and observed systems occupy different regions of parameter space. Our simulations have $\dot M \approx (6.7$--$9.5) \times 10^{19}$~g~s$^{-1}$ and $B_\star \approx (3$--$10) \times 10^{10}$~G, whereas the \texttt{bwcycl} fits give $\dot M \sim10^{18}$~g~s$^{-1}$ and $B\sim 10^{12}$--$10^{13}$~G. 

The mean lateral luminosities from our simulations, $\overline{L_{\rm side}} \approx (2.0$--$4.5) \times 10^{38}$~erg~s$^{-1}$, are comparable to the observed outburst luminosities, however $L_{\rm side}$ is not an observable parameter. The isotropic (apparent) luminosities $L_{\rm iso}$ evaluated at $d=500\,r_{\rm g}$, are reported in Figure~12 of \citet{Kayanikhoo2025a} and range from approximately $39$ to $84\,L_{\rm Edd}$. These values, along with those of \cite{Zhang2025} lie in the ULX range, one to two orders of magnitude above the isotropic-equivalent $L_{\rm X}$ of the observed outbursts as shown in Fig.~\ref{fig:hmax}.

The models also differ in their geometry. BW07 assumes a one-dimensional cylindrical column with a prescribed velocity profile. The radiation-MHD simulations of \citet{Zhang2025} use a split-monopole magnetic field and magnetic opacity calculations. In contrast, our simulations follow accretion from the disk onto the NS through a dipolar magnetic field, allowing the column geometry to develop self-consistently. In addition, the models treat radiation processes differently. 

In summary, our GRRMHD simulations show that across four models, the accretion column becomes narrower with strength of the stellar magnetic field, although the accretion rate also increases. Despite the different accretion rates, magnetic fields, and geometries of the models, the area-equivalent column radii overlap with the upper range of the values inferred from spectral model BW07. However, the characteristic column heights remain considerably smaller than those from the spectral model. The apparent luminosity from the simulations in the range of 40--80~$L_{\rm Edd}$ is substantially higher than that inferred from the spectral model. 

We will perform simulations with various accretion rates and initial conditions; particularly, simulations with lower $\dot M$ and higher $B_\star$ will test whether the agreement in column radius still holds and whether the difference in column height can be reduced. In simulations with stronger magnetic fields, we will implement magnetic opacity. 

\section*{Acknowledgments}
We thank the development team of \textsc{Koral}, Olek S\k{a}dowski and David Abarca. FK acknowledges the program Vouchers for Universities in the Moravian-Silesian Region (registration number CZ.10.03.01/00/23\_042/0000390), within the project Accreting Magnetised neutron stars. This work was supported by the Ministry of Education, Youth and Sports of the Czech Republic through the e-INFRA CZ (ID:90254). We gratefully acknowledge Poland's high-performance Infrastructure PLGrid ACC Cyfronet AGH for providing computer facilities and support within computational grant no. PLG/2025/018773. MK acknowledges full funding by the EU HORIZON- MSCA-2023-DN Project 101168906 “TALES: Time-domain Analysis to study the Life-cycle and Evolution of Super-massive black holes.”
\bibliography{references}
\end{document}